\documentclass[twoside,twocolumn,9pt]{article}
\usepackage{extsizes}
\usepackage[super,sort&compress,comma]{natbib} 
\usepackage[version=3]{mhchem}
\usepackage[left=1.5cm, right=1.5cm, top=1.785cm, bottom=2.0cm]{geometry}
\usepackage{balance}
\usepackage{mathptmx}
\usepackage{amsmath}
\usepackage{sectsty}
\usepackage{graphicx} 
\usepackage{lastpage}
\usepackage[format=plain,justification=justified,singlelinecheck=false,font={stretch=1.125,small,sf},labelfont=bf,labelsep=space]{caption}
\usepackage{float}
\usepackage{fancyhdr}
\usepackage{fnpos}
\usepackage[english]{babel}
\addto{\captionsenglish}{%
	
}
\usepackage{array}
\usepackage{droidsans}
\usepackage{charter}
\usepackage[T1]{fontenc}
\usepackage[usenames,dvipsnames]{xcolor}
\usepackage{setspace}
\usepackage[compact]{titlesec}
\usepackage{hyperref}
\usepackage{ulem}

\definecolor{cream}{RGB}{222,217,201}

\newcommand{\beginsupplement}{%
        \setcounter{table}{0}
        \renewcommand{\thetable}{S\arabic{table}}%
        \setcounter{figure}{0}
        \renewcommand{\thefigure}{S\arabic{figure}}%
        \setcounter{equation}{0}
        \renewcommand{\theequation}{S\arabic{equation}}
     }

\begin{document}

\pagestyle{fancy}
\thispagestyle{plain}
\fancypagestyle{plain}{
	%%%HEADER%%%
	\renewcommand{\headrulewidth}{0pt}
}
%%%END OF HEADER%%%

%%%PAGE SETUP - Please do not change any commands within this section%%%
\makeFNbottom
\makeatletter
\renewcommand\LARGE{\@setfontsize\LARGE{15pt}{17}}
\renewcommand\Large{\@setfontsize\Large{12pt}{14}}
\renewcommand\large{\@setfontsize\large{10pt}{12}}
\renewcommand\footnotesize{\@setfontsize\footnotesize{7pt}{10}}
\makeatother

\renewcommand{\thefootnote}{\fnsymbol{footnote}}
\renewcommand\footnoterule{\vspace*{1pt}% 
	\color{cream}\hrule width 3.5in height 0.4pt \color{black}\vspace*{5pt}} 
\setcounter{secnumdepth}{5}

\makeatletter 
\renewcommand\@biblabel[1]{#1}            
\renewcommand\@makefntext[1]% 
{\noindent\makebox[0pt][r]{\@thefnmark\,}#1}
\makeatother 
\renewcommand{\figurename}{\small{Fig.}~}
\sectionfont{\sffamily\Large}
\subsectionfont{\normalsize}
\subsubsectionfont{\bf}
\setstretch{1.125} %In particular, please do not alter this line.
\setlength{\skip\footins}{0.8cm}
\setlength{\footnotesep}{0.25cm}
\setlength{\jot}{10pt}
\titlespacing*{\section}{0pt}{4pt}{4pt}
\titlespacing*{\subsection}{0pt}{15pt}{1pt}
%%%END OF PAGE SETUP%%%

%%%FOOTER%%%
\fancyfoot{}
\fancyfoot[LO,RE]{\vspace{-7.1pt}\includegraphics[height=9pt]{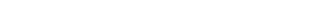}}
\fancyfoot[CO]{\vspace{-7.1pt}\hspace{13.2cm}\includegraphics{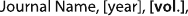}}
\fancyfoot[CE]{\vspace{-7.2pt}\hspace{-14.2cm}\includegraphics{head_foot/RF}}
\fancyfoot[RO]{\footnotesize{\sffamily{1--\pageref{LastPage} ~\textbar  \hspace{2pt}\thepage}}}
\fancyfoot[LE]{\footnotesize{\sffamily{\thepage~\textbar\hspace{3.45cm} 1--\pageref{LastPage}}}}
\fancyhead{}
\renewcommand{\headrulewidth}{0pt} 
\renewcommand{\footrulewidth}{0pt}
\setlength{\arrayrulewidth}{1pt}
\setlength{\columnsep}{6.5mm}
\setlength\bibsep{1pt}
%%%END OF FOOTER%%%

%%%FIGURE SETUP - please do not change any commands within this section%%%
\makeatletter 
\newlength{\figrulesep} 
\setlength{\figrulesep}{0.5\textfloatsep} 

\newcommand{\topfigrule}{\vspace*{-1pt}% 
	\noindent{\color{cream}\rule[-\figrulesep]{\columnwidth}{1.5pt}} }

\newcommand{\botfigrule}{\vspace*{-2pt}% 
	\noindent{\color{cream}\rule[\figrulesep]{\columnwidth}{1.5pt}} }

\newcommand{\dblfigrule}{\vspace*{-1pt}% 
	\noindent{\color{cream}\rule[-\figrulesep]{\textwidth}{1.5pt}} }

\makeatother
%%%END OF FIGURE SETUP%%%

%%%TITLE, AUTHORS AND ABSTRACT%%%
\twocolumn[
\begin{@twocolumnfalse}
	{\includegraphics[height=30pt]{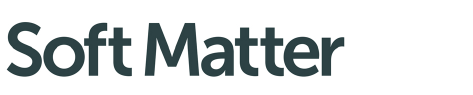}\hfill\raisebox{0pt}[0pt][0pt]{\includegraphics[height=55pt]{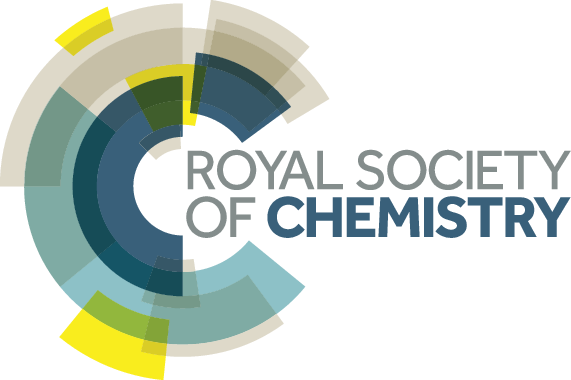}}\\[1ex]
		\includegraphics[width=18.5cm]{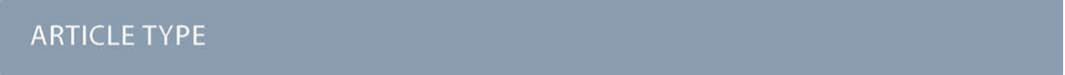}}\par
	\vspace{1em}
	\sffamily
	\begin{tabular}{m{4.5cm} p{13.5cm} }
		
		\includegraphics{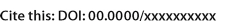} & \noindent\LARGE{\textbf{Destructive effect of fluctuations on the performance of a Brownian gyrator $^\dag$} }\\%Article title goes here instead of the text "This is the title"
		\vspace{0.3cm} & \vspace{0.3cm} \\

		& \noindent\large{Pascal Viot,\textit{$^{a\ddag}$} Aykut Argun,\textit{$^{b\ddag}$}, 
		Giovanni Volpe,	\textit{$^{b\ddag}$},Alberto Imparato  \textit{$^{c\ddag}$}, Lamberto Rondoni \textit{$^{d\ddag}$},  and Gleb Oshanin\textit{$^{a\ddag}$}} \\%Author names go here instead of "Full name", etc.
		
		\includegraphics{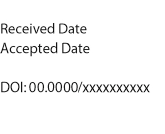} & \noindent\normalsize{
			The Brownian gyrator (BG) is  often called a minimal model of a nano-engine performing a rotational motion,
			judging solely upon the fact that  
			in non-equilibrium conditions its torque, 
			angular momentum ${\cal L}$ and angular velocity $\cal W$ have non-zero mean values. 
			For a time-discretized  model, which is most adapted for the analysis of an essentially discrete-time data garnered in experiments or numerical simulations,  we calculate the
			previously unknown probability density functions (PDFs)
			of ${\cal L}$ and $\cal W$. 
%			We find that when the time-step $\delta t \to 0$, both PDFs converge to 
%			uniform distributions 
%			with diverging 
%			variances.
			For finite time-step $\delta t$, the PDF of ${\cal L}$ has exponential tails and all moments are therefore well-defined, 
			but  the noise-to-signal ratio can attain big values for small $\delta t$. 
			Conversely, the PDF of ${\cal W}$ exhibits heavy power-law tails and its mean ${\cal W}$ is the only existing moment. 
			The BG is therefore not an engine in the common sense:  it does not exhibit regular rotations on each run
			and  its fluctuations are not only a minor nuisance --  on contrary, their effect is completely destructive for the performance.   
			Our theoretical predictions are confirmed by numerical simulations and experimental data.
			We discuss some plausible improvements 
			of the model which may result in a more systematic behavior.
} \\%The abstrast goes here instead of the text "The abstract should be..."
		
	\end{tabular}
	
\end{@twocolumnfalse} \vspace{0.6cm}

]
%%%END OF TITLE, AUTHORS AND ABSTRACT%%%

%%%FOOTNOTES%%%

\footnotetext{\textit{$^{a}$~Sorbonne Universit\'e, CNRS, Laboratoire de Physique Th\'eorique de la Mati\`ere Condens\'ee (UMR CNRS 7600), 4 Place Jussieu, 75252 Paris Cedex 05, France. Email: pascal.viot@sorbonne-universite.fr}}
\footnotetext{\textit{$^{b}$~Physics Department, University of Gothenburg, 412 96 Gothenburg Sweden}}
	
\footnotetext{\textit{$^{c}$~Department of Physics and Astronomy, University of Aarhus, Ny 
		Munkegade, Building 1520, DK--8000 Aarhus C, Denmark}}

\footnotetext{\textit{$^{d}$~Dipartimento di Scienze Matematiche, Politecnico di Torino, Corso Duca degli Abruzzi 24,  10129 Torino, Italy }}

%Please use \dag to cite the ESI in the main text of the article.
%If you article does not have ESI please remove the the \dag symbol from the title and the footnotetext below.
\footnotetext{\dag~Electronic Supplementary Information (ESI) available: [details of any supplementary information available should be included here]. See DOI: 00.0000/00000000.}
%additional addresses can be cited as above using the lower-case letters, c, d, e... If all authors are from the same address, no letter is required

\footnotetext{\ddag~`These authors contributed equally to this work.}

%%%END OF FOOTNOTES%%%

%%%FONT SETUP - please do not change any commands within this section
\renewcommand*\rmdefault{bch}\normalfont\upshape
\rmfamily

\section{Introduction.} The fundamental working principles of macroscopic thermal engines, 
producing deterministic translational or rotational motion of large objects, are well-established in 
classical thermodynamics. 
At the same time, the latter 
does not explain workings of different kinds of microscopic motors, encountered in molecular or cellular biology, as well as in other rather diverse biophysical systems. Such tiny machines operate while effectively coupled to a single heat bath (or even several heat baths at the same time), making the concepts of macroscopic classical thermodynamics inapplicable.
Nonetheless they 
work and are sufficiently efficient, e.g., to
control transport of organelles throughout the cell and to produce
 rotary motion of bacterial flagella. 

Considerable progress in understanding general concepts  and various aspects of the performance of microscopic motors,  
as well as necessary and sufficient conditions for their work has been achieved during the last decades \cite{reimann,berg,haenggi,sekimoto,seifert,volpe,sergio}. In parallel, various
case-by-case analyses elucidated precise mechanisms
to rectify fluctuations produced by thermal baths and to convert their energy into useful work.
However, most of the available analyses concern
the average characteristics of the performance of molecular motors, such as \textit{mean} velocities or \textit{mean} rotational frequencies. Still little is known about the magnitude of fluctuations around these values and their impact on the performance of microscopic motors.

The Brownian gyrator (BG) is  often called a minimal stochastic model of a microscopic heat engine that performs, on average, a rotational motion along closed orbits in non-equilibrium conditions. It consists, 
as realized experimentally\cite{argun}, of an optically-trapped\cite{optical}
Brownian colloidal particle that is simultaneously coupled to two heat baths kept at different 
temperatures $T_x$ and $T_y$, 
acting along perpendicular directions and maintaining a non-equilibrium steady-state. An alternative experimental realization of the BG  using equivalent  electric circuits has been also devised\cite{alberto,alberto2}.
First theoretically introduced  for the analysis of effective temperatures in non-equilibrium systems \cite{peliti}, 
this model was subsequently revisited\cite{rei} arguing  that  a \textit{systematic} torque on the particle is generated once $T_x \neq T_y$. To support this claim, it was shown that indeed the mean torque is non-zero in such non-equilibrium conditions.  Further analyses 
 evaluated the mean angular velocity \cite{dotsenko,pascal,bae}, and examined
  various aspects of the dynamical and steady-state behaviors for delta-correlated in time noises \cite{alberto,alberto2,crisanti,lahiri,lamberto,tyagi,Chang2021,NEXUS,viale,Dotsenko2022,Dotsenko2023}, for the BG in the quantum regime \cite{Fogedby2018}, for the noises with long-ranged temporal correlations \cite{Nascimento2021,Squarcini2022a},  as well as the effects of inertia\cite{bae,pastur} on the performance of the BG, of anisotropy of fluctuations\cite{Miangolarra2021,Miangolarra2022}, and the statistics of the entropy production\cite{mazzolo}.   
  
 The BG operates in heat baths and evidently the 
  torque, angular momentum ${\cal L}$ and angular velocity ${\cal W}$
 exhibit sample-to-sample fluctuations and also fluctuate during each given realization. 
 The spread of fluctuations and their typical values are unknown, hence their actual 
 impact on the BG performance remains elusive.
 In this regard, it is unclear to which extent, e.g.,  the torque can be called \textit{systematic} \cite{rei}, 
 such that the question whether the BG can be indeed called a "motor"  remains unanswered.
 
 In the quest for the answer,  
 we study here the probability density functions (PDFs) of  ${\cal L}$ and ${\cal W}$ for a  time-discretized  BG model. We proceed to show that both PDFs are effectively broad such that the spread of fluctuations around mean values of ${\cal L}$ and ${\cal W}$ is much bigger than these mean values themselves. 
This implies that non-zero values  of first moments only indicate some trend in an ensemble of BGs, but no systematic behavior can be observed for individual realizations. Our findings signify that the BG cannot be called a \textit{motor} in the common sense, and we suggest some improvements which may result in a more regular behavior.
Our 
  theoretical predictions are confirmed by numerical simulations and also appear consistent with experimental data \cite{argun}. 
   
\section{The Model.} The model consists 
of two Langevin equations for two linearly coupled Ornstein-Uhlenbeck processes,  
each living at its own temperature, $T_x$ or $T_y$ (both measured in units of the Boltzmann constant). 
In its simplest settings \cite{peliti,rei}, the BG model is defined by 
\begin{equation}
\begin{split}
\label{a}
\dot{X}_t = - X_t + u Y_t + \xi_x(t) \,, \\
\dot{Y}_t = - Y_t + u X_t + \xi_y(t) \,, 
\end{split}
\end{equation}
with $|u| < 1$,  and $\xi_x(t)$ and $\xi_y(t)$ being independent 
Gaussian noises with zero mean and covariance function
\begin{equation}
\label{noises}
\overline{\xi_{i}(t) \xi_{j}(t')} = 2  T_j \delta_{i,j} \delta(t-t') \,, \quad i,j = x,y\,. 
\end{equation}
In eqs. \eqref{noises}, the overbar denotes averaging over realizations of noises, while $\delta_{i,j}$ and $\delta(t)$ are the Kronecker symbol and the delta-function, respectively. For $u=0$, eqs. \eqref{a} describe two independent Ornstein-Uhlenbeck processes.  The solution of eqs. \eqref{a} and the position PDF are presented in the Electronic Supplementary Inrormation (ESI).  The solution for arbitrary coefficients is also available \cite{alberto,alberto2,sara}.

We focus here on the following two random variables:\\ 
-- the first is the \textit{magnitude} ${\cal L}$ of the angular momentum, 
 \begin{equation}
 \label{L}
 {\cal L} =  X_t \dot{Y}_t - Y_t \dot{X}_t \,.
 \end{equation}
-- the second is the \textit{magnitude} ${\cal W}$ of the angular velocity,
\begin{equation}
\label{W}
{\cal W} = \frac{X_t \dot{Y}_t - Y_t \dot{X}_t}{X^2_t+Y^2_t} \,,
\end{equation} 
where  
the term in the denominator is the moment of inertia ${\cal I} = X^2_t+Y^2_t$. In both Eqs. \eqref{L} and \eqref{W} we assumed that the mass $m=1$.
For the standard discrete-time (with time-step $\delta t$) analogue of eqs. \eqref{a} (see eqs. \eqref{az} in the ESI), 
 we present below exact expressions for the PDFs of the random variables ${\cal L}$ and ${\cal W}$.   We also note parenthetically that the properties of the time-integrated ${\cal L}$, which can be interpreted as an area enclosed by the stochastic trajectory $X_t,Y_t$, have been studied\cite{duBuisson2022,duBuisson2023} recently for the BG model.

\begin{figure}
\centering
\includegraphics[width=0.96\columnwidth]{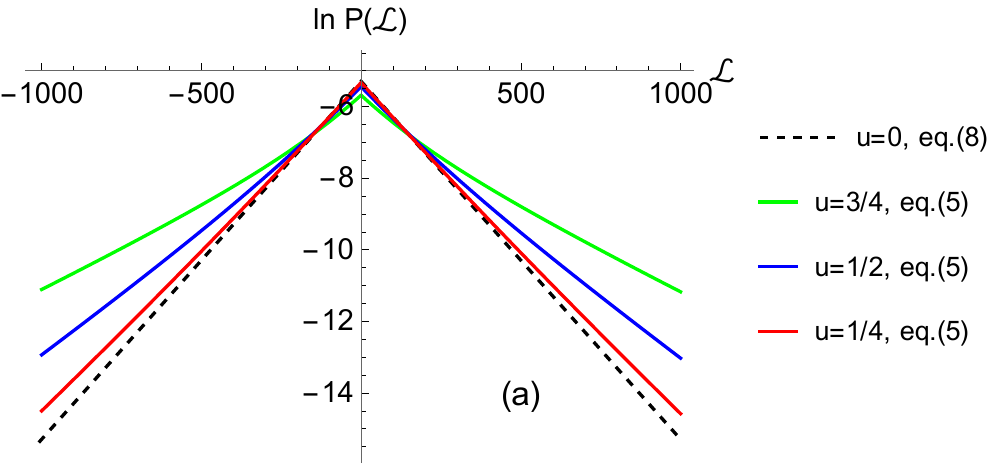}
\includegraphics[width=0.96\columnwidth]{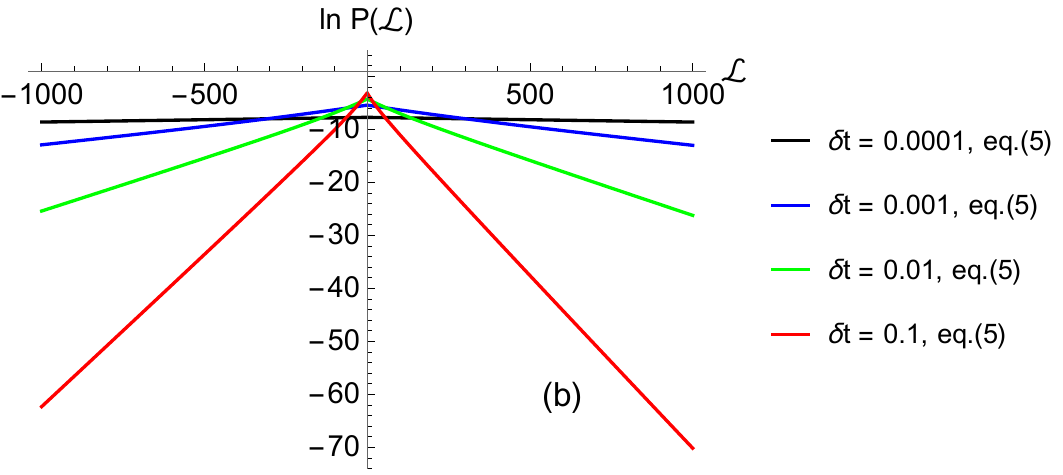}
\caption{(color online) Angular momentum in the steady-state.  
$\ln P({\cal L})$ versus ${\cal L}$ for $T_x = 1$ and $T_y = 5$. Panel (a): $\delta t = 0.001$ 
and variable $u$.  Dashed black curve  is the decoupled form  in eq.~\eqref{u00}, the solid curves - a numeric evaluation of the integral in eq.~\eqref{probL}. 
%with $u = 1/4$ (red),  
%$u=1/2$ (blue) and $u=3/4$ (green). 
Panel (b): $u=1/2$ and variable $\delta t$. 
Comparison with the
numerical simulations results is presented in the ESI.
}  
\label{F:1}
\end{figure}

\section{ Probability density function of the angular momentum.}  As shown
 in the ESI, the PDF of the magnitude ${\cal L}$ of the angular momentum 
 is given by 
\begin{equation}
\begin{split}
\label{probL}
P({\cal L}) &= \overline{\delta\Big({\cal L} - \left(X_t \dot{Y}_t - Y_t \dot{X}_t\right)\Big)}\\
&=\frac{1}{4 \pi d} \int^{2 \pi}_{0} \frac{d\theta}{\Lambda(\theta)} \exp\left(-\Xi(\theta) \, |{\cal L}| \right)  \,,
\end{split}
\end{equation}
where $d =  \sqrt{4 a b - c^2}$ with $a$, $b$ and $c$ defined in the ESI, 
\begin{equation}
\label{Xi}
\Xi(\theta) = \frac{\left(\Lambda(\theta) - {\rm sign}({\cal L}) u \cos(2 \theta) \right)}{2 \left(T_y \cos^2(\theta) + T_x \sin^2(\theta)\right)} \, \delta t \,,
\end{equation}
and
\begin{equation}
\begin{split}
\label{Lambda}
\Lambda(\theta) &= \Big(u^2 \cos^2(2 \theta) + \frac{4 \left(T_y \cos^2(\theta) + T_x \sin^2(\theta)\right)}{d^2 \delta t} \\ &\times \left(b \cos^2(\theta) + a \sin^2(\theta) + c \cos(\theta) \sin(\theta) \right) \Big)^{1/2} \,.
\end{split}
\end{equation}
Except for the trivial case $u=0$, when $P({\cal L})$ obeys
\begin{equation}
\label{u00}
P_{u=0}({\cal L}) = \sqrt{\frac{\delta t}{8 T_x T_y}} \exp\left(- \sqrt{\frac{\delta t}{2 T_x T_y}} \, |{\cal L}| \right) \,,
\end{equation}
the integral in eq. \eqref{probL} cannot be performed exactly and we evaluate it numerically. 
 In Fig. \ref{F:1} we plot $P({\cal L})$  in eq. \eqref{probL} as function of ${\cal L}$ 
for unequal temperatures $T_x$ and $T_y$, four values of the coupling parameter $u$  at fixed small $\delta t $ (panel (a)), and three values of $\delta t$ at fixed $u$ (panel (b)). 
This plot together with an asymptotic analysis permits us to make the following general statements: \\
-- $P({\cal L})$ is peaked at ${\cal L} = 0$, i.e., for most of trajectories one should observe ${\cal L}=0$. Hence, a non-zero value 
\begin{equation}
\begin{split}
\label{cum1}
&\overline{\cal L} = u (T_x - T_y) \, , 
\end{split}
\end{equation}
of the first moment is supported by atypical trajectories, and stems from the asymmetry (see below) of the PDF.  The first moment exists in the limit $\delta t \to 0$, which agrees with previous results. The variance of ${\cal L}$ obeys
\begin{equation}
\begin{split}
\label{cum2}
{\rm Var}({\cal L}) &=  2 u^2 (T_x - T_y)^2 \\&+ \frac{(1 + u^2 \delta t) (4 T_x T_y + u^2 (T_x - T_y)^2) }{(1 - u^2) \delta t}\,.
\end{split}
\end{equation}
As expected, for the overdamped model the variance, as well as all higher cumulants \textit{diverge
} in the continuous-time limit $\delta t \to 0$, 
 meaning that the angular momentum is not self-averaging. 
However, for finite $\delta t$, as it is the case in experimental or numerical analyses, all moments remain finite, although
  the observed scatter of values of ${\cal  L}$ calculated from different realizations of the process can be very large.\\
-- For finite $\delta t$, the large-$L$ tails of the PDF follow
\begin{equation}
\label{as}
P({\cal L}) \simeq \exp\left(- |{\cal L}|/L_{\pm}\right) \,,
\end{equation}
where the symbol $\simeq$ denotes the leading in the large-${\cal L}$ limit behavior,
$L_{+}$ corresponds to the right (${\cal L} > 0$) and $L_{-}$ - to the left (${\cal L} < 0$) tails of the PDF. Mathematically, $L_{\pm}$ are solutions of the fourth-order algebraic equation that defines the maximal values of the function $1/\Xi(\theta)$ and therefore are rather complicated functions of the system's parameters. In general, $L_{\pm}$ are increasing functions of the coupling $u$ (see Fig. \ref{F:1}(a)).\\ 
-- When $\delta t \to 0$, both $L_{\pm}$ diverge, because $L_{\pm} \simeq 1/\sqrt{\delta t}$.  Hence,  the PDF becomes some ${\cal L}$-independent constant of order $O(\sqrt{\delta t})$ in the interval ${\cal L} \in [-1/\sqrt{\delta t},1/\sqrt{\delta t}]$,  
\begin{equation}
\label{PLas}
P({\cal L}) \simeq \frac{\sqrt{2 (1 - u^2) \delta t}}{4 \pi (T_x+T_y)} f(u,\Delta) \,, \quad \Delta = \frac{T_x - T_y}{T_x + T_y} \,,
\end{equation}
where $f(u,\Delta)$ is defined in the ESI.
That being, $P({\cal L})$ tends to a uniform distribution on this interval of values of ${\cal L}$. 
Such a behavior is apparent in  Fig. \ref{F:1}(b).

\section{Probability density function of the angular velocity.} 
\begin{figure}
\centering
\includegraphics[width=0.96\columnwidth]{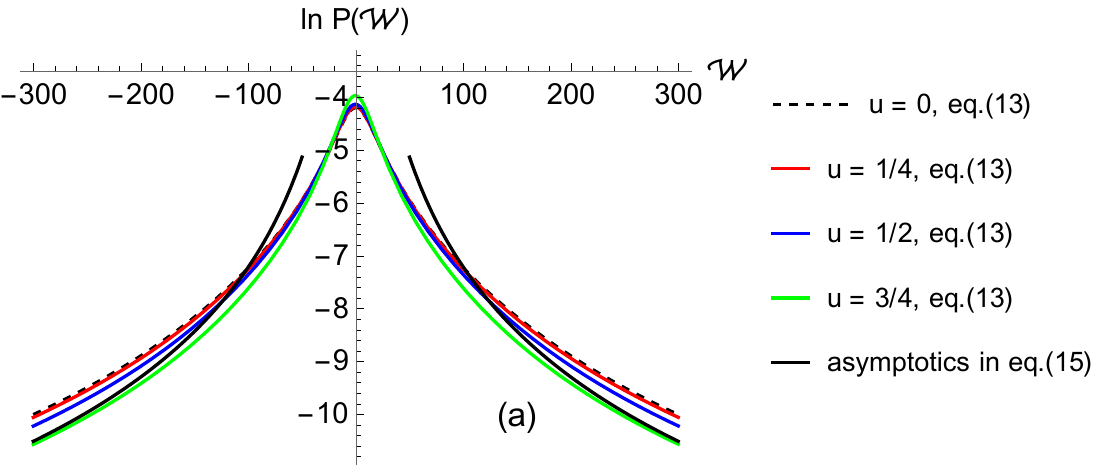}
\includegraphics[width=0.96\columnwidth]{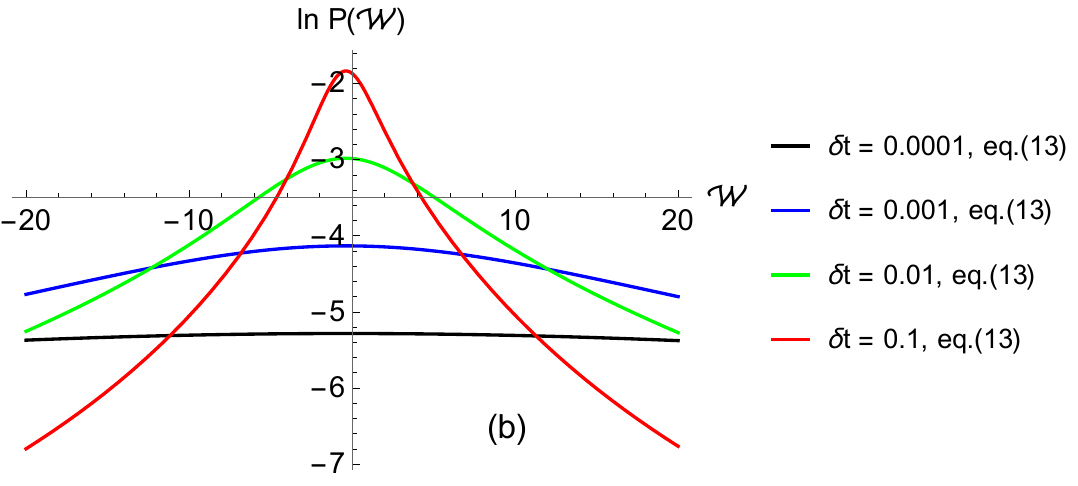}
\caption{(color online) Angular velocity in the steady-state. $\ln P({\cal W})$ in eq. \eqref{probW} vs ${\cal W}$ for $T_x = 1$ and $T_y = 5$. Panel (a): $\delta t = 10^{-3}$ and variable $u$.  Panel (b):  $u=1/2$ and variable $\delta t$.  Comparison with numerics is presented in the ESI. }  
\label{F:2}
\end{figure}

The derivation of $P({\cal W})$ is presented in the ESI. It is defined as the following integral
\begin{equation}
\begin{split}
\label{probW}
&P({\cal W}) = \overline{\delta\left({\cal W} - \left(X_t \dot{Y}_t - Y_t \dot{X}_t\right)/\left(X^2_t+Y^2_t\right)\right)}\\
&= \frac{1}{2 \pi d \delta t} \int^{2 \pi}_0 \frac{(T_y \cos^2(\theta) + T_x \sin^2(\theta)) \, d\theta}{\left({\cal W}^2 - 2 u \cos(2 \theta) {\cal W} + \Lambda^2(\theta)\right)^{3/2}}  \,,
\end{split}
\end{equation}
with $\Lambda(\theta)$ being defined in eq. \eqref{Lambda}.
For $u =0$ the integral in eq. \eqref{probW} can be performed exactly (see the ESI), 
while in the general case it is amenable only to a numerical and asymptotic analyses. 
The PDF in eq. \eqref{probW} is depicted  in Fig. \ref{F:2} as function of ${\cal W}$ for unequal temperatures, several values of the coupling parameter and several values of $\delta t$. On the basis of the numerical plots and asymptotic analysis, the following conclusions can be drawn :\\
-- The maximum of $P({\cal W})$ displaces from zero for $u \neq 0$ and is rounded, in contrast to the cusp-like behavior of $P({\cal L})$ at ${\cal L}=0$. Hence, $P({\cal W})$ is an analytic function in the vicinity of the most probable value.\\
-- The first moment of the PDF $P({\cal W})$ obeys (see the ESI)
\begin{equation}
\label{meanW}
\overline{\cal W} =  \frac{u \sqrt{1-u^2} (T_x - T_y)}{\sqrt{4 T_xT_y + u^2 (T_x - T_y)^2}}  \,,
\end{equation}
which is a well-known result for the BG evolving in continuous time ($\delta t \to 0$).  We note that the difference between
the most probable ${\cal W}$ and $\overline{\cal W}$ is much less pronounced than the one observed for ${\cal L}$. In particular, for $T_x =1$, $T_y = 5$, $u=1/2$ and $\delta t = 1$, we have that the most probable ${\cal W}  \approx - 0.38$, while $\overline{\cal W}  \approx - 0.35$.  \\
-- For finite $\delta t$, the large-${\cal W}$ tails of the PDF obey
\begin{equation}
\label{as1}
P({\cal W}) \simeq \frac{\sqrt{1 - u^2} (T_x + T_y)}{\sqrt{4 T_x T_y + u^2 (T_x-T_y)^2} \delta t} \frac{1}{|{\cal W}|^3} \,,
\end{equation}   
where the symbol $\simeq$ signifies that we deal with the leading in this  limit behavior.   The salient feature is that, for finite $\delta t$, the PDF is characterized by heavy power-law tails such that $\overline{\cal W}$ in eq. \eqref{meanW} is the \textit{only} existing moment - in other words, already the variance of ${\cal W}$ is infinitely large.   
Therefore, $\overline{\cal W}$ does not have much of a physical significance but merely indicates some non-zero trend 
 in a statistical ensemble of Brownian gyrators, when $T_x \neq T_y$. For a single BG it is therefore absolutely unlikely to observe 
 a regular rotation, i.e., a motor-like behavior. The asymptotic form in 
 eq. \eqref{as1} for $u=3/4$ is depicted in Fig. \ref{F:2}(a) (solid black line) and we observe that 
 it converges to $P({\cal W})$ in eq. \eqref{probW} for ${\cal W} \approx 100$.
 Subdominant correction term to the asymptotic form in eq. \eqref{as1} is displayed in the ESI.    \\
 -- When $\delta t \to 0$, $P({\cal W})$ converges to a uniform distribution on 
 a growing (when $\delta t \to 0$) interval $[-1/\sqrt{\delta t},1/\sqrt{\delta t}]$, 
 \begin{equation}
 P({\cal W}) \simeq \frac{(4 T_x T_y + u^2 (T_x - T_y)^2)}{8 \pi (T_x + T_y)^2} \sqrt{2 (1 - u^2) \delta t} g(u,\Delta) \,, 
\label{PWas}
 \end{equation}
 where $g(u,\Delta)$ in presented in the ESI.  In Fig. \ref{F:2}(b) we depict $P({\cal W})$ for progressively smaller values of $\delta t$, highlighting a transition to a uniform distribution.

%\vspace{0.1in} 
 
\section{Comparison with experimental results.}  
To perform a comparison with experiments, we employed the data acquired by Argun et al.  
\cite{argun}, who experimentally realized the BG model by placing  
a Brownian particle in an elliptical optical potential and simultaneously maintaining it in contact with two heat baths kept at different temperatures. This was achieved by using a single optically-trapped \cite{volpe} colloidal particle (polystyrene particle of diameter $1.98\, {\rm \mu m}$) suspended in an aqueous solution at room temperature ($292$ K). The ellipticity of the potential was produced by altering the laser beam's intensity profile via a spatial light modulator. The isotropic thermal environment was made anisotropic via a fluctuating electric field with a near-white frequency spectrum applied along the $x$-direction. This electric fluctuating field was generated by two thin wires with electric white noise placed on either side of the optical trap and raised the bath noise temperature along the $x$-direction to 1750 K, while the bath noise temperature along the $y$-direction remained at 292 K (room temperature).
In this way, unequal temperatures along the $x$- and $y$-directions were produced to establish a non-equilibrium steady-state.

\begin{figure}
\centering
\includegraphics[width=0.46\columnwidth]{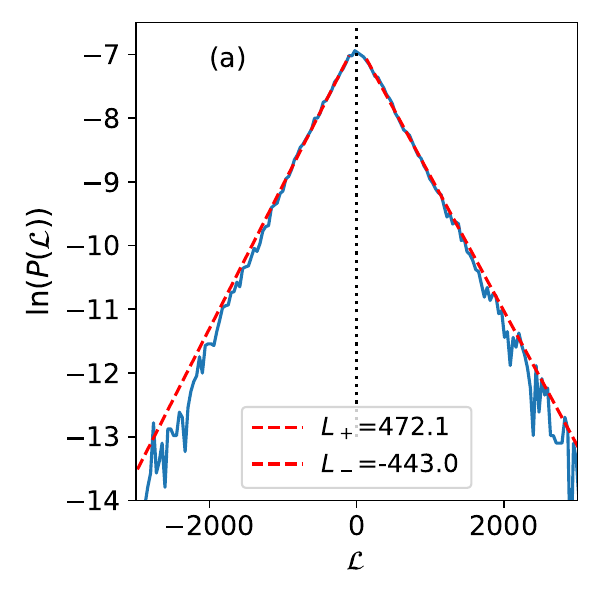}
\includegraphics[width=0.46\columnwidth]{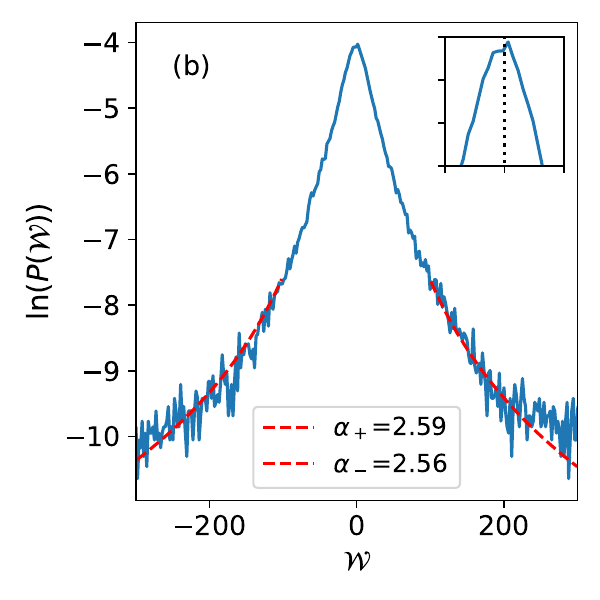}
\caption{(color online) Comparison with experimental data. Noisy blue curves - the PDFs $P({\cal L})$ (panel  (a)) and $P({\cal W})$ (panel (b))
deduced from experimental  data with two unequal temperatures, dashed red curves represent the fit of $P({\cal L})$ by an exponential function in eq. \eqref{as} (panel (a)) and of $P({\cal W})$ by power-laws: $1/{\cal W}^{\alpha_+}$ for ${\cal W} > 0$ and $1/|{\cal W}|^{\alpha_-}$ for ${\cal W} < 0$ (panel (b)). The inset shows a small displacement of the maximum from ${\cal W}=0$ as evidenced by experimental data.}  
\label{F:3}
\end{figure}

In Fig. \ref{F:3} we depict $P({\cal L})$ and $P({\cal W})$ 
evaluated from a statistical ensemble of 
discrete-time BG trajectories recorded in the above experimental set-up. We observe 
that for $P({\cal L})$ the predicted form in eq. \eqref{as} is fully consistent with the experimental data; the tails 
are exponential and  $P({\cal L})$ shows a cusp-like behavior at
the most probable value ${\cal L}=0$.  Also our predictions that $P({\cal W})$ is centered around a non-zero 
most probable value of ${\cal W}$ and that the large-${\cal W}$ decay is a power-law appear to be valid features of the BG 
model. However,  we note a slight discrepancy between the predicted value of the exponent ($= 3$),  
and the values $\alpha_{\pm}$ deduced from the experimental data, which appear to be somewhat smaller and hence, the tails appear somewhat heavier. It seems rather  paradoxical at the first glance that the distribution displays heavy power-law tails and that, at the same time the mean value can be obtained accurately from the data. The reason is the following: the angular velocity is obtained by computing the difference of angles between two consecutive positions, which is bounded between $-\pi$ and $\pi$. Consequently, for a finite time-step the variance remains finite and the mean value can be therefore deduced. As discussed in ESI, a
truncation of the distribution is likely to modify its shape close to the boundaries. Numerical simulations support our claim (see the ESI).

%\vspace{0.1in}

\section{Discussion.} 
 To recapitulate,  we revisited  a well-studied minimalistic model of a molecular motor -- the so-called Brownian gyrator -- which is believed to perform a rotational motion in non-equilibrium conditions.  While previous works concentrated  exclusively on the analysis of the \textit{mean} values of the characteristics of the gyration process (torque, angular momentum ${\cal L}$ and angular velocity ${\cal W}$), here we looked on this model from a broader perspective 
and calculated exact probability density functions of these random variables for a discrete-time BG model with a time-step $\delta t$. This permitted us to obtain a fully comprehensive picture of the BG
performance. 

We showed that $\delta t$ is a crucial parameter. For $\delta t  \to 0$, i.e., in the continuous-time limit, both PDFs tend to a uniform distribution with a vanishing amplitude and diverging variance.  For fixed $\delta t$, the PDF  of ${\cal L}$ 
has exponential tails which ensures the existence of all moments.
 However, the variance of ${\cal L}$ appears to be much larger than the squared first moment, meaning that here the \textit{noise} is always bigger than the \textit{signal}.   In other words, the value of the angular momentum varies significantly  from one realization  of Gaussian noises to another. More strikingly, the PDF of the angular velocity has heavy power-law tails and, in fact, the mean angular velocity is the only existing moment. 
 In conclusion, fluctuations play a very destructive role on the performance of the BG, such that it cannot be considered as
 a common sense motor for which only some small amount of noise is permitted. On the contrary, for the BG the 
 noise prevails producing a very erratic behavior. While the existence of first moments indicates some 
 trend in an ensemble of such Brownian gyrators, no systematic rotation will take place on the level of 
 a single realization. Indeed, only a non-vanishing fraction of realizations needs to 
 have this property for the whole statistic to be affected.

Our theoretical predictions are confirmed by numerical simulations and are also consistent with the data drawn from the experimental realization\cite{argun} 
of the BG model. We note parenthetically that although heavy-tailed PDFs are 
common in probability theory,
 the list of their occurrences in realistic physical systems is rather short.  Our prediction for the PDF  of the angular velocity in conjunction with the experimental data\cite{argun} adds a  valuable example to this list.

Finally, we note that minimal models often play an important role in the analysis of the behavior of complex systems. 
 While they often take into account only some basic features discarding a majority of side-processes,
  they are nonetheless very instructive  because they provide conceptual insights helping the understanding of the dominant mechanisms. This is the reason why the BG model has attracted such an attention in the past.  
 In view of our results, the model as it stands does not appear to be really performant. 
 Therefore, a legitimate question is how the model can be improved, still remaining at a minimalistic level, to show a better performance. 
 
 One of possible issues is the definition of noise. We observed here that
in the limit $\delta t \to 0$, the PDFs become progressively more defocused 
which is clearly detrimental for the performance. Consequently, replacing a continuous-time white-noise by 
a discrete noise with a bounded amplitude may be beneficial (and is definitively more appropriate for describing experimental systems). 
We note, as well, that in the limit $\delta t \to \infty$ 
	the model becomes deterministic and the rotational motion disappears, which suggests that there may be some optimal 
	discretization of the noise. 
	
Second, we remark that the
power-law tails of the probability density function of the angular velocity stem from
realizations of trajectories in which the BG spends most of time in the immediate vicinity of the origin. 
Such realizations, which have a big statistical weight, produce anomalously high values 
of the moment of inertia (see the ESI for the PDF of the latter), and therefore, 
anomalously high positive or negative values of the angular velocity (see eq. \eqref{W}). 
% Inspecting the definition
%of the angular velocity, eq. \eqref{W},  one may conclude that such realizations correspond to  trajectories with non-zero 
%(positive or negative) values of the angular momentum and very small values of the moment of inertia ${\cal  I}$. In other words, 
%high values of the angular velocity occur when the BG circulates along closed orbits that concentrate around the origin.  
%The probability density function of the moment of inertia is presented in the ESI and we observe that actually ${\cal I}=0$ 
%is the most probable value. 
Therefore, a physically plausible suggestion to devise a motor with a more regular behavior is 
to forbid the BG to enter a finite-radius ball centered at the origin. This can be realized in 
practice by designing a potential with a strong repulsion from the origin at short scales, and an attraction at larger 
scales, such that the closed orbits along which the BG circulates will have a finite length. Here, again some optimization with respect to the range of attractive/repulsive interactions may take place. While having a too short range of repulsive interactions is detrimental, as we have shown here, a too long range will also be detrimental for the performance 
because the closed orbits will be too long and hence, an angular velocity too small. 

Lastly, the model may apparently be improved to some extent along the lines suggested in \cite{bae}, i.e., by introducing inertial terms into eqs. \eqref{a} which will bound the acceleration.

%\vspace{0.1in}

\section*{Acknowledgements}
The authors wish to thank Olivier B\'enichou, Martin-Luc Rosinberg and Luca Peliti for helpful discussions.
This research was performed under
the auspices of Italian National Group of Mathematical Physics (GNFM) of INdAM.

\pagebreak

\onecolumn
\section{Electronic Supplementary information}
\beginsupplement

\subsection{A. Position probability density function} 

The solutions of eqs. \eqref{a} with the initial conditions $X_0 = Y_0 = 0$ and $\dot{X}_0 = \dot{Y}_0 = 0$ are explicitly given by
\begin{equation}
\begin{split}
\label{b}
X_t &= e^{-t} \int^t_0 d\tau \, e^{\tau} \, {\rm cosh}\left(u (t- \tau)\right) \, \xi_x(\tau) + e^{-t} \int^t_0 d\tau \, e^{\tau} \, {\rm sinh}\left(u (t- \tau)\right) \, \xi_y(\tau) \,, \\
Y_t &= e^{-t} \int^t_0 d\tau \, e^{\tau} \, {\rm sinh}\left(u (t- \tau)\right) \, \xi_x(\tau) + e^{-t} \int^t_0 d\tau \, e^{\tau} \, {\rm cosh}\left(u (t- \tau)\right) \, \xi_y(\tau) \,.
\end{split}
\end{equation}
We  focus on the characteristic function of the form
\begin{equation}
\phi(\nu_1,\nu_2) = \overline{\exp\Big(i \nu_1 X_t + i \nu_2 Y_t\Big)} \,. 
\end{equation} 
The averaging in the latter expression can be performed straightforwardly to give
\begin{equation}
\phi(\nu_1,\nu_2) =  \exp\left(- T_x \int^t_0 d\tau \, Q_x^2(t,\tau) - T_y \int^t_0 d\tau \, Q_y^2(t,\tau) \right) \,,
\end{equation}
where 
\begin{equation}
\begin{split}
\label{Q}
Q_x(t,\tau) & = e^{-(t - \tau)} \Big(\nu_1 \, {\rm cosh} \left(u (t -\tau)\right)   + \nu_2 \, {\rm sinh} \left(u (t -\tau)\right) \Big) \,,\\
Q_y(t,\tau) & = e^{-(t - \tau)} \Big(\nu_1 \, {\rm sinh} \left(u (t -\tau)\right)   + \nu_2 \, {\rm cosh} \left(u (t -\tau)\right)  \Big) \,.
\end{split}
\end{equation}
Performing the integrals in eqs. \eqref{Q}, we find
\begin{equation}
\begin{split}
\label{q}
&T_x \int^t_0 d\tau \, Q_x^2(t,\tau) + T_y \int^t_0 d\tau \, Q_y^2(t,\tau) = a \nu_1^2 +  b \nu_2^2 - c \nu_1 \nu_2 \,,
\end{split}
\end{equation}
where we have used the shortenings
\begin{equation}
\begin{split}
\label{aa}
a = \frac{T_x p + T_y q}{4 (1-u^2)} \,,\quad 
b= \frac{T_x q + T_y p}{4 (1-u^2)} \,, \quad
c = \frac{(T_x+T_y) \, l}{2 (1-u^2)} \,, 
\end{split}
\end{equation}
with
\begin{equation}
\begin{split}
\label{p}
p &= 2 - u^2 - e^{-2 t} \Big({\rm cosh}\left(2 u t\right) + u \,  {\rm sinh}\left(2 u t\right)\Big) - (1-u^2) e^{- 2 t} \,, \\
q &= u^2 - e^{-2 t} \Big({\rm cosh}\left(2 u t\right) + u \,  {\rm sinh}\left(2 u t\right)\Big) + (1-u^2) e^{- 2 t} \,, \\
l & = u - e^{-2 t} \Big(u \, {\rm cosh}\left(2 u t\right) +  {\rm sinh}\left(2 u t\right)\Big) \,.
\end{split}
\end{equation}
Having defined the characteristic function, we find eventually the position probability density function
by merely inverting the Fourier transform:
\begin{equation}
\begin{split}
\label{P}
P(X_t,Y_t)  &= \frac{1}{(2 \pi)^2} \int^{\infty}_{-\infty} \int^{\infty}_{-\infty} d\nu_1 \, d\nu_2 \, e^{- i \nu_1 X_t - i \nu_2 Y_t} \, \phi(\nu_1,\nu_2) \\
&= \frac{1}{2 \pi d} \exp\left(- \frac{b X_t^2 + a Y_t^2 + c X_t Y_t}{d^2}\right) \,,
\end{split}
\end{equation} 
where $d = \sqrt{4 a b - c^2}$ and the coefficients $a$, $b$ and $c$ are defined in eqs. \eqref{aa} and \eqref{p}. 

\subsection{B. Probability density function of the angular momentum} 

\subsubsection{1. Characteristic function $\Phi_{\cal L}(\nu)$.}

In virtue of eqs. \eqref{a}, the magnitude of angular momentum can be cast into the form 
\begin{equation}
\begin{split}
\label{def}
{\cal L}&= X_t \dot{Y}_t - \dot{X}_t Y_t \equiv  u \left(X_t^2 - Y_t^2\right) + X_t \xi_y(t) - Y_t \xi_x(t) \,,
\end{split}
\end{equation}
such that its characteristic function  is formally defined by
\begin{equation}
\begin{split}
\label{c}
\Phi_{\cal L}(\nu) &= \overline{\exp\left(i \nu {\cal L}\right)} \\
&= \overline{\exp\Big(i \nu u \left(X_t^2 - Y_t^2\right) + i \nu X_t \xi_y(t) - i \nu Y_t \xi_x(t) \Big)} \,.
\end{split}
\end{equation}
We note now that the values of the noise variables $\xi_x(t)$ and $\xi_y(t)$ are statistically decoupled 
from the values of the positions $Y_t$ and $X_t$ at this very time moment, such that, in principle, we can average
over the instantaneous values of the noises. The problem is that the values of the noises are ill-defined. Taking this into account, 
and also having in mind a comparison with numerical and experimental data, 
which actually correspond to trajectories recorded at discrete time moments, 
we turn to time-discretized version of the continuous-time Langevin equations 
\eqref{a}. In discrete-time with  time-step $\delta t$, a conventional representation of eqs. \eqref{a} is 
\begin{equation}
\begin{split}
\label{az}
\frac{X_{t+\delta t} - X_t}{\delta t} = - X_t + u Y_t + \sqrt{\frac{2 T_x}{\delta t}} \, \eta_x(t) \,, \\
\frac{Y_{t+\delta t} - Y_t}{\delta t}  = - Y_t + u X_t +  \sqrt{\frac{2 T_y}{\delta t}} \, \eta_y(t) \,, 
\end{split}
\end{equation}
where now the noises $\eta_{x,y}(t)$ are dimensionless, independent, normally-distributed random variables with zero
mean and unit variance, drawn independently at each discrete time step. In terms of eqs. \eqref{az}, the magnitude of the angular momentum takes the form
\begin{equation}
\label{Lz}
{\cal L} = u \left(X_t^2 - Y_t^2\right) + \sqrt{\frac{2 T_y}{\delta t}} \, X_t \, \eta_y(t)  -  \sqrt{\frac{2 T_x}{\delta t}} \, Y_t  \,\eta_x(t) \,,
\end{equation}
and therefore the characteristic function of the angular momentum reads
\begin{equation}
\begin{split}
\label{cc}
\Phi_{\cal L}(\nu)  
= \overline{\exp\left(i \nu u \left(X_t^2 - Y_t^2\right) + i \nu  \sqrt{\frac{2 T_y}{\delta t}} \, X_t \eta_y(t) - i \nu  \sqrt{\frac{2 T_x}{\delta t}}  \, Y_t \eta_x(t) \right)} \,.
\end{split}
\end{equation}
Performing the averaging over the variables $\eta_x(t) $ and $\eta_y(t)$, we readily find that $\Phi_{\cal L}(\nu)$ obeys
\begin{equation}
\begin{split}
\label{c1}
\Phi_{\cal L}(\nu) &= \overline{\exp\left(- \left(\frac{T_y}{\delta t} \, \nu^2 - i \nu u \right) X_t^2 - \left(\frac{T_x}{\delta t} \, \nu^2 + i \nu u \right) Y_t^2 \right)}  \,,
\end{split}
\end{equation}
which yields, upon averaging with the PDF in eq. \eqref{P},  the following explicit result:
\begin{equation}
\begin{split}
\label{Phigen}
\Phi_{\cal L}(\nu) &= \Bigg(1 - 4 i (a - b) u \, \nu + 4 \left(\frac{b T_x + a T_y}{\delta t} + u^2 d^2\right) \, \nu^2 - 4 i u d^2 \frac{\left(T_x - T_y\right)}{\delta t} \, \nu^3 + 4 d^2 \frac{T_x T_y}{\delta t^2}  \, \nu^4
\Bigg)^{-1/2} \,,
\end{split}
\end{equation}
where the coefficients $a$, $b$ and $c$ are defined in eqs. \eqref{aa} and \eqref{p}, and $d = \sqrt{4 a b - c^2}$.  Therefore, the characteristic function 
$\Phi_{\cal L}(\nu)$ is simply the inverse of a square-root of a quartic polynomial of $\nu$  and becomes ill-defined, as mentioned above, 
 in the limit $\delta t \to 0$.
In the limit $t \to \infty$, the expression \eqref{Phigen} attains the form
\begin{equation}
\begin{split}
\label{phi1}
\Phi_{\cal L}(\nu) &= \Bigg(1 - 2 i u \, (T_x- T_y) \, \nu + \frac{(1+u^2 \delta t)\left(4 T_x T_y + u^2 (T_x-T_y)^2\right)}{(1 - u^2) \delta t} \, \nu^2 \\
&- i \, \frac{u (T_x- T_y) \left(4 T_x T_y + u^2 (T_x-T_y)^2\right)}{(1 - u^2) \delta t} \, \nu^3 + \frac{T_x T_y \left(4 T_x T_y + u^2 (T_x-T_y)^2\right)}{(1 - u^2) \delta t^2} \, \nu^4\Bigg)^{-1/2} \,.
\end{split}
\end{equation}
The above expression permits us to access the usual properties characterizing the probability density function, i.e., the moments and the cumulants. 
In particular, differentiating eq. \eqref{phi1} once and twice, and setting  $\nu=0$ afterwards, we find explicit expressions 
for the first moment
and the variance of ${\cal L}$ in eqs. \eqref{cum1} and  \eqref{cum2}.

\subsubsection{2. Probability density function $P({\cal L}) $.}

The probability density function of the angular momentum is given by
\begin{equation}
\begin{split}
\label{zz}
P({\cal L}) &= \frac{1}{2 \pi} \int^{\infty}_{-\infty} \int^{\infty}_{-\infty} dX_t \, dY_t  \, P(X_t,Y_t)  \int^{\infty}_{-\infty} d\nu \, \exp\Big(- i \nu {\cal L} - \left(\frac{T_y}{\delta t} \, \nu^2 - i \nu u \right) X_t^2 - \left(\frac{T_x}{\delta t} \, \nu^2 + i \nu u \right) Y_t^2 \Big) \\
&=  \sqrt{\frac{\delta t}{4 \pi}} \int^{\infty}_{-\infty} \int^{\infty}_{-\infty} \frac{dX_t \, dY_t }{\sqrt{T_y X_t^2 + T_x Y_t^2}}  P(X_t,Y_t) \, \exp\left(- \frac{\delta t \left({\cal L} - u \left(X_t^2 - Y_t^2\right) \right)^2}{4 \left(T_y X_t^2 + T_x Y_t^2\right)}\right) \,. 
\end{split}
\end{equation}

\begin{figure}
\centering
\includegraphics[width=0.3\columnwidth]{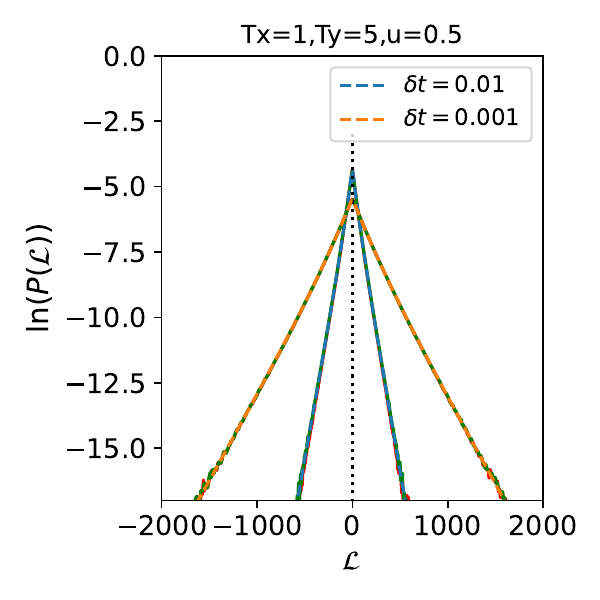}
\includegraphics[width=0.3\columnwidth]{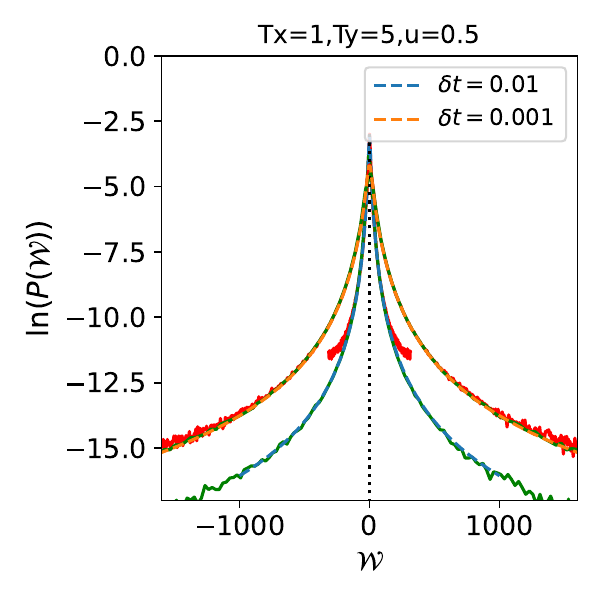}
\includegraphics[width=0.3\columnwidth]{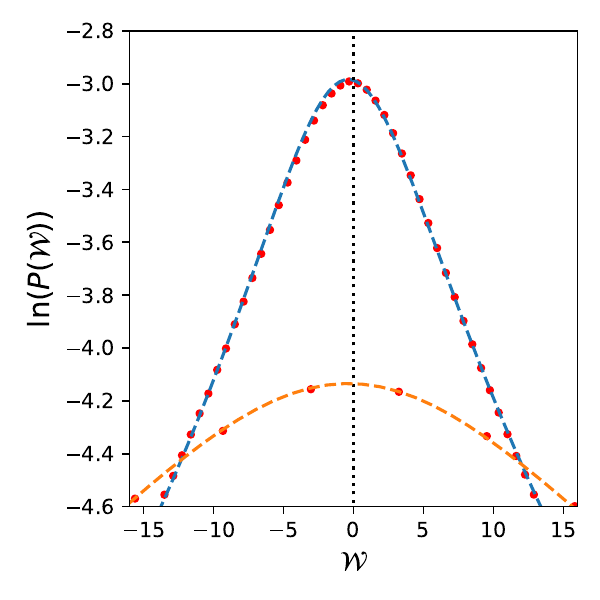}
\caption{Comparison of theoretical predictions for the probability density functions in eqs. \eqref{probL} and \eqref{probW} and the
  results of numerical simulations for two values of the time-step $\delta t$: $\delta t = 0.01$ (blue) and $\delta t = 0.001$ (magenta); and fixed $T_x=1$, $T_y=5$ and $u=0.5$. 
 Left panel: Logarithm of $P({\cal  L})$ versus ${\cal L}$. Dashed curves correspond to the exact expression \eqref{probL}, whereas the noisy curves depict the simulation results obtained using the  approaches I and II  (see below). Central panel:  Logarithm of $P({\cal W})$ versus ${\cal W}$. Dashed  curves depict the exact expression \eqref{probW}, and the noisy curves - the simulation results.
  Right panel: Zoom of the region around the maximal value of $P({\cal W})$. Dashed curves are theoretical predictions in eq. \eqref{probW}, 
while the symbols indicate the results of numerical simulations.  }
\label{FIGSM2}
\end{figure}

Turning to polar coordinates through the transformation $X_t = \rho \cos(\theta)$ and $Y_t = \rho \sin(\theta)$, we formally rewrite the expression in the second line in eq. \eqref{zz} as
\begin{equation}
\begin{split}
\label{proba}
P({\cal L}) &= \frac{1}{2 d} \sqrt{\frac{\delta t}{\pi}} \int^{2 \pi}_{0} \frac{d\theta}{\sqrt{T_y \cos^2(\theta) + T_x \sin^2(\theta)}}  \int^{\infty}_{0}  d\rho \,  \exp\left( - \frac{\delta t \left({\cal L} - u \rho^2 \cos(2 \theta) \right)^2}{4 \rho^2 (T_y \cos^2(\theta) + T_x \sin^2(\theta))} \right)\\
& \times \exp\left(- \left(\frac{b \cos^2(\theta) + a \sin^2(\theta) + c \cos(\theta) \sin(\theta)}{d^2}\right) \rho^2 \right).
\end{split}
\end{equation}
Performing the integral over $\rho$, we find the expression \eqref{probL}.
In Fig. \ref{FIGSM2} (left panel) we present a comparison between our theoretical prediction for $P({\cal L})$ in eq. \eqref{probL} and the results of numerical simulations. One observes a perfect agreement. 

\subsubsection{3. Finite $\delta t$. Large-${\cal L}$ asymptotic behavior of $P({\cal L}) $.}

The function $\Xi(\theta)$, eq. \eqref{Xi}, is an oscillatory function of the polar angle $\theta$, which has two minima (and two maxima) of equal
depth (height) on the interval $[0,2 \pi]$ (see Fig. S2). In general, the minima corresponding to ${\cal L} > 0$ and ${\cal L} < 0$ are attained at somewhat different values of $\theta$ and have somewhat different depths, which signifies that $P({\cal L})$ is asymmetric around ${\cal L}=0$. 
In the limit ${\cal L} \to \infty$, the integral in eq. \eqref{probL} is entirely dominated by the  behavior of $\Xi$ in the close vicinity 
of the minima, which yields the exponential asymptotic form in eq. \eqref{as}. Differentiating $\Xi(\theta)$ with respect to $\theta$ and equating the resulting expression to zero, we find an equation that determines the positions of the extrema of $\Xi$. 
%This is a fourth-order algebraic equation, which is unsolvable for arbitrary values of parameters. We therefore consider only the limiting case when the coupling parameter $u$ is close (but not equal) to zero and $\delta t$ is also small, in which limit 
%it is possible to derive explicit forms of $L_{\pm}$. 
This is a fourth-order algebraic equation whose solution has a rather cumbersome form. We therefore consider only the limiting case when the coupling parameter $u$ is close (but not equal) to zero and $\delta t$ is also
small, in which limit it is possible to present rather compact explicit results. 
A rather cumbersome but straightforward analysis shows that here 
\begin{equation}
\label{T}
\frac{1}{L_{\pm}} = \sqrt{\frac{\delta t}{2 T_x T_y}} \left(1 - \frac{(T_y + T_x)}{2 \sqrt{T_x T_y}} |u| + O\left(u^2\right)\right) \,.
\end{equation}
When $u=0$, we recover from the latter expression the result in eq. \eqref{u00}.

\begin{figure}
\centering
\includegraphics[width=0.3\columnwidth]{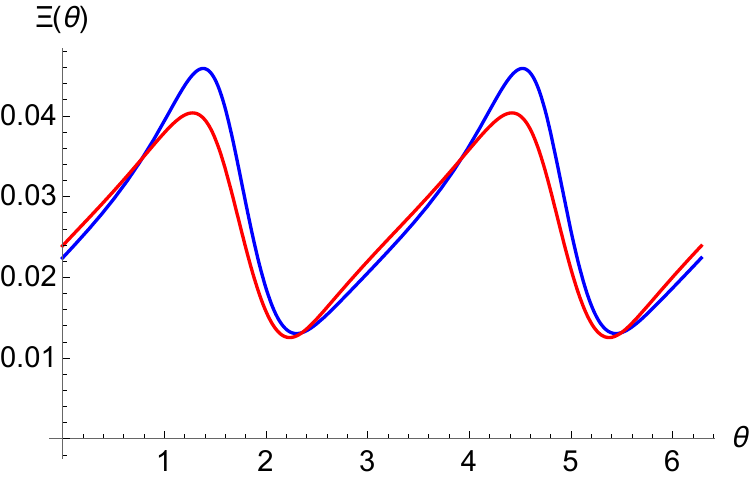}
\caption{The function $\Xi(\theta)$ in eq. \eqref{Xi} versus the polar angle $\theta$ for $T_x = 1$, $T_y = 5$, $u=1/2$ and $\delta t = 0.1$. Red solid curve corresponds to ${\cal L} > 0$, while the blue one - to ${\cal L} < 0$.  }
\label{F:6}
\end{figure}

Another situation, in which a systematic analysis of the large-${\cal L}$ asymptotic behavior is possible, 
is that of thermodynamic equilibrium, i.e., $T_x=T_y = T$. Here, we pursue a slightly different approach focusing on the characteristic function in eq. \eqref{phi1}, which becomes (for $u^2 \delta t \ll 1)$
\begin{equation}
\begin{split}
\label{phi4}
\Phi_{\cal L}(\nu) &= \Bigg(1  + \frac{4 T^2}{(1 - u^2) \delta t} \, \nu^2  + \frac{4 T^4}{(1 - u^2) \delta t^2} \, \nu^4\Bigg)^{-1/2} \,.
\end{split}
\end{equation}
The latter expression can be formally rewritten as
\begin{equation}
\begin{split}
\Phi_{\cal L}(\nu) & = \frac{\sqrt{(1 - u^2)}}{\sqrt{\left(1 + \dfrac{2 T^2 \nu^2}{\delta t}\right)^2 - u^2}} \\&= \sqrt{(1 - u^2)} \int^{\infty}_0  dx \, I_0\left(|u| \, x\right) \, \exp\left(- \left(1 + \frac{2 T^2 \nu^2}{\delta t}\right) x\right) \,,
\end{split}
\end{equation}
and hence, the probability density function attains the form
\begin{equation}
\label{t}
P({\cal L}) = \frac{\sqrt{(1 - u^2) \delta t}}{2 \sqrt{2 \pi} T} \int^{\infty}_0 \frac{dx}{\sqrt{x}} \, I_0\left(|u| \, x\right) \, \exp\left(- x -  \frac{\delta t {\cal L}^2}{8 T^2 x}\right) \,,
\end{equation}
where $I_0$ is the modified Bessel function of the first kind.
Setting $u=0$ in eq. \eqref{t}, we recover our eq. \eqref{u00} with $T_x=T_y=T$. 
Upon some inspection, we realize that the large-${\cal L}$ (more precisely, $\delta t {\cal L}^2/(8 T^2)$ is to be large and $u$ bounded away from zero) 
behavior of $P({\cal L})$ is supported by the large-$x$ behavior of the integrand. To this end, we take advantage of the Hankel's asymptotic expansion of the modified Bessel function
\begin{equation}
I_0(z) = \frac{e^z}{\sqrt{2 \pi z}} \left(1 + \frac{1}{8 z} + \frac{9}{2 (8 z)^2} + O\left(\frac{1}{z^3}\right)\right) \,.
\end{equation} 
Inserting the latter expansion into eq. \eqref{t} and performing integrations, 
we arrive at the converging 
asymptotic large-${\cal L}$ expansion of the form
\begin{equation}
\begin{split} 
P({\cal L}) &= \left(\frac{1-u^2}{8 \pi T |u \cal L|}\right)^{1/2} \left(\frac{2 \delta t}{(1 - |u|)}\right)^{1/4} \exp\left(- \sqrt{\frac{(1 - |u|) \delta t}{2}} \frac{|{\cal L}|}{T}\right) \\
&\times \Bigg(1 - \frac{ (3 |u| - 2)T}{4 \sqrt{2 (1 - |u|) \delta t} \, |u {\cal L}|}
+ \frac{3 T^2}{16 \delta u^2 {\cal L}^2} + O\left(\frac{1}{|{\cal L}|^3}\right)
\Bigg) \,.
\end{split}
\end{equation}
Consequently, we recover the asymptotic form in eq. \eqref{as} with 
\begin{equation}
L_+=L_-= \sqrt{\frac{2}{(1- |u|) \delta t}}  \, T \,,
\end{equation}
which is fully consistent with eq. \eqref{T}. Note that the higher order terms in $\delta t$ and $u$ are evidently missing here
because we assumed that $u^2 \delta t \ll 1$.

\subsubsection{4. Continuous-time limit $\delta t \to 0$. Asymptotic behavior of $P({\cal L}) $.}

We concentrate on the behavior of the probability density function $P({\cal L}) $ in the limit $\delta t \to 0$. 
One notices that in this limit the second term in eq. \eqref{Lambda} becomes much larger than the first term, (i.e., $u^2 \cos^2(2 \theta)$), which can be therefore safely neglected. In consequence, we have
\begin{equation}
\label{exp}
\Lambda(\theta) \simeq \frac{2}{d \sqrt{\delta t}} \left(\left(T_x \sin^2(\theta) + T_y \cos^2(\theta)\right) \left(b \cos^2(\theta) + a \sin^2(\theta) + c \cos(\theta) \sin(\theta)\right) 
\right)^{1/2} \,,
\end{equation} 
which implies that $\Lambda(\theta)$ diverges when $\delta t \to 0$. In this limit, the first term in the nominator in eq. \eqref{Xi} is much larger than the second term, (i.e., ${\rm sign}({\cal L}) u \cos(2 \theta)$), which can be neglected. Therefore, the function  $\Xi(\theta)$ in eq. \eqref{Xi} obeys (in this limit)
\begin{equation}
\begin{split}
\Xi(\theta) & \simeq \frac{\Lambda(\theta)}{2 \left(T_y \cos^2(\theta) + T_x \sin^2(\theta)\right)} \delta t \\
&= \left(\frac{b \cos^2(\theta) + a \sin^2(\theta) + c \cos(\theta) \sin(\theta) }{T_x \sin^2(\theta) + T_y \cos^2(\theta)}\right)^{1/2} \frac{\sqrt{\delta t}}{d}
\end{split}
\end{equation}
and hence, vanishes in proportion to the square-root of $\delta t$. 

\begin{figure}
\centering
\includegraphics[width=0.3\columnwidth]{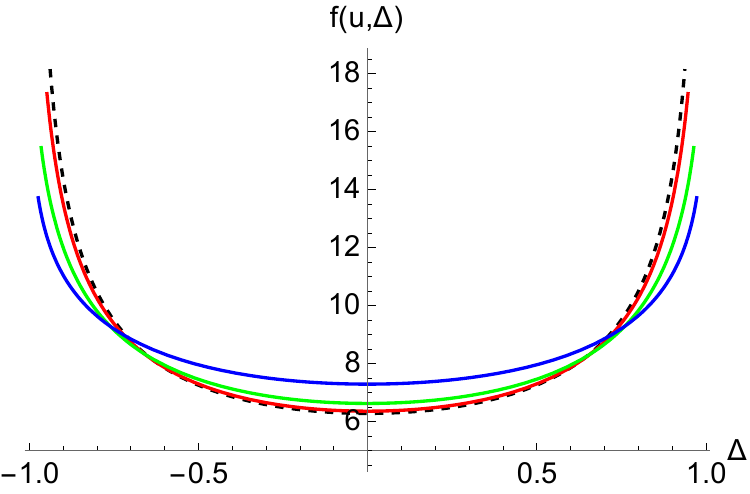}
\includegraphics[width=0.3\columnwidth]{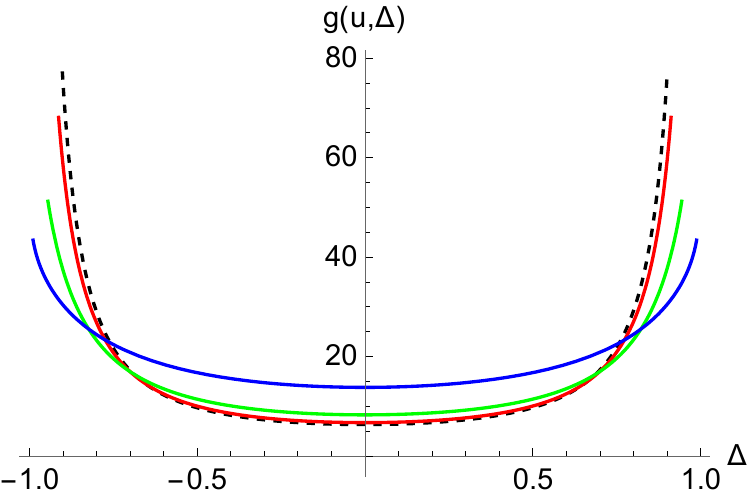}
\caption{Small $\delta t$ limit. Left panel : $f(u,\delta)$ in eq. \eqref{f} as function of $\Delta = (T_x - T_y)/(T_x+T_y)$. Right panel:  $g(u,\delta)$ in eq. \eqref{g} as function of $\Delta$. In both panels the dashed curve corresponds to
$u = 0$, red solid curve - to $u=1/4$, green - to $u = 1/2$  and blue - to $u = 3/4$.  }
\label{F:5}
\end{figure}

We assume that $t \to \infty$, in which case the coefficients $a$, $b$ and $c$ attain sufficiently simple time-independent forms. 
Using the definitions of the coefficients $a$, $b$ and $c$, and conveniently rewriting 
the terms entering the expression \eqref{exp} as
\begin{equation}
\begin{split}
\label{expr}
& T_x \sin^2(\theta) + T_y \cos^2(\theta) = \frac{T_x + T_y}{2} \left(1 - \Delta \cos(2 \theta)\right) \,, \quad \Delta = \frac{T_x - T_y}{T_x + T_y} \,, \\
& b \cos^2(\theta) + a \sin^2(\theta) + c \cos(\theta) \sin(\theta)  = \frac{T_x + T_y}{4 (1 - u^2)} \left(1 + u \sin(2 \theta) - (1-u^2) \Delta \cos(2 \theta) \right) \,.
\end{split}
\end{equation} 
Further on, we suppose that ${\cal L}$ is finite (i.e., ${\cal L} \delta t$ is small when $\delta t \to 0$) and expand the expression in the right-hand-side of eq. \eqref{probL} into the Taylor series in powers of $\delta t$.
In doing so, we realize that, in the leading in the limit $\delta t$ order, the probability density function $P({\cal L})$ in eq. \eqref{probL}
is given by eq. \eqref{PLas} with
\begin{equation}
\label{f}
f(u,\Delta) = \int^{2 \pi}_0 \frac{d\phi}{ \sqrt{\left(1 - \Delta \cos(\phi)\right)  \left(1 + u \sin(\phi) - (1-u^2) \Delta \cos(\phi) \right)}} \,.
\end{equation}
The integral in the latter equation cannot be performed exactly. To get an idea of the behavior of $f(u,\Delta)$, we depict it in Fig. \ref{F:5} as function of $\Delta$ for several values of the coupling parameter $u$. Note that $f(u,\Delta)$ is (logarithmically) diverging in the limit $\Delta \to \pm 1$, i.e., when either of the temperatures vanishes. It was shown
in \cite{lamberto} that this limit is rather peculiar and the case when either of the temperatures vanishes is to be treated separately. We therefore suppose that both $T_x,T_y > 0$. 
Further on, correction terms to the leading asymptotic behavior in eq. \eqref{PLas} are of order ${\cal L} \delta t$. We verified that 
the coefficient in front of ${\cal L} \delta t$ is finite, such that a uniform distribution of ${\cal L}$ in the limit $\delta t \to 0$ is a valid feature.

\subsection{C. Probability density function of the angular velocity.}

\subsubsection{1. Characteristic function $\Phi_{\cal W}(\nu)$.}

Using the discretized Langevin eqs. \eqref{a} (see eqs. \eqref{az} above), 
we rewrite the angular velocity as 
\begin{equation}
\begin{split}
\label{def2}
{\cal W}&=  \frac{u \left(X_t^2 - Y_t^2\right)}{X_t^2+ Y_t^2} + \frac{\sqrt{2 T_y} \, X_t \, \eta_y(t)  -  \sqrt{2 T_x} \, Y_t  \,\eta_x(t) }{\sqrt{\delta t} \left(X_t^2+ Y_t^2\right)} \,,
\end{split}
\end{equation}
such that  characteristic function $\Phi_{\cal W}(\nu) $ is given by :
\begin{equation}
\Phi_{\cal W}(\nu) = \overline{\exp\left(i \nu u \frac{\left(X_t^2 - Y_t^2\right)}{X_t^2+ Y_t^2} + i \nu \frac{\sqrt{2 T_y} \, X_t \, \eta_y(t)  -  \sqrt{2 T_x} \, Y_t  \,\eta_x(t) }{\sqrt{\delta t} \left(X_t^2+ Y_t^2\right)}\right)} \,.
\end{equation}
Again, noticing that the instantaneous values of the noises $\eta_x(t)$ and $\eta_y(t)$ at time $t$ have no effect on the instantaneous positions $X_t$ and $Y_t$ at this very time, we straightforwardly average over $\eta_x(t)$ and $\eta_y(t)$ to get
\begin{equation}
\begin{split}
\label{c2}
\Phi_{\cal W}(\nu) &= \overline{\exp\left(i \nu u \frac{\left(X_t^2 - Y_t^2\right)}{X_t^2+ Y_t^2}     - \frac{T_y \, \nu^2 X_t^2}{\delta t (X_t^2 + Y_t^2)^2} - \frac{T_x \, \nu^2 Y_t^2}{\delta t (X_t^2 + Y_t^2)^2} \right)}  \,.
\end{split}
\end{equation}

\subsubsection{2. Probability density function $P({\cal W})$.}

The probability density function $P({\cal W})$ of the angular velocity is formally defined as
\begin{equation}
\begin{split}
P({\cal W}) &= \frac{1}{2 \pi} \int^{\infty}_{-\infty} d\nu \, \Phi_{\cal W}(\nu) \, e^{-I \nu {\cal W}} \\
&=  \frac{1}{2 \pi} \int^{\infty}_{-\infty} d\nu \, e^{-I \nu {\cal W}} \int^{\infty}_{-\infty} \int^{\infty}_{-\infty} dX_t \,
 dY_t  \, P(X_t,Y_t) \\ & \times \, \exp\left(i \nu u \frac{\left(X_t^2 - Y_t^2\right)}{X_t^2+ Y_t^2}     - \frac{T_y \, \nu^2 X_t^2}{\delta t (X_t^2 + Y_t^2)^2} - \frac{T_x \, \nu^2 Y_t^2}{\delta t (X_t^2 + Y_t^2)^2} \right) \,.
 \end{split}
\end{equation}
Changing the integration variables to polar coordinates, we formally  rewrite the latter expression as
\begin{equation}
\begin{split}
\label{fin}
P({\cal W}) &= \frac{1}{(2 \pi)^2 \sqrt{4 a b - c^2}} \int^{2 \pi}_0 d\theta \int^{\infty}_{-\infty} d\nu \, \exp\Big(i \nu \left(u \cos(2 \theta) - {\cal W} \right)\Big) \\
& \times \int^{\infty}_0 \rho \, d\rho \, \exp\left(- \frac{\nu^2 \left(T_y \cos^2(\theta) + T_x \sin^2(\theta)\right)}{\delta t \rho^2} - \left(\frac{b \cos^2(\theta) + a \sin^2(\theta) + c \cos(\theta) \sin(\theta)}{4 a b - c^2}\right) \, \rho^2\right) \\
&= \frac{1}{\pi^2 \delta t \sqrt{4 a b - c^2}} \int^{2 \pi}_0 \frac{(T_y \cos^2(\theta) + T_x \sin^2(\theta)) \, d\theta}{\sqrt{\Lambda^2(\theta) - u^2 \cos^2(2 \theta)}}  \int^{\infty}_{0} \nu \, d\nu \, \cos\Big(\nu \left(u \cos(2 \theta) - {\cal W} \right)\Big) \, \\&\times  K_1\left(\sqrt{\Lambda^2(\theta) - u^2 \cos^2(2 \theta)} \, \nu\right) \,, 
 \end{split}
\end{equation}
where $K_1$ is the modified Bessel function and $\Lambda(\theta)$ is defined in eq. \eqref{Lambda}. Performing the integral over $\nu$, we arrive at our result in eq. \eqref{probW}.

\subsubsection{3. Finite $\delta t$. Large-${\cal W}$ asymptotic behavior of $P({\cal W}) $.}

Asymptotic large-${\cal W}$ behavior of the probability density function $P({\cal W})$ in eq. \eqref{probW} can be accessed very directly by merely expanding the denominator in inverse powers of ${\cal W}$. Verifying that all the integrals defining the coefficients in this expansion exist, we find
\begin{equation}
\label{fin0}
P({\cal W}) = \frac{(T_x + T_y)}{2 d \delta t} \frac{1}{|{\cal W}|^3} - \frac{3 u (T_x - T_y)}{4 d \delta t} \frac{1}{{\cal W}^4} + o\left(\frac{1}{{\cal W}^4}\right) \,.
\end{equation}
In the limit $t \to \infty$, eq. \eqref{fin0} gives
\begin{equation}
\label{fin1}
P({\cal W}) = \frac{\sqrt{1 - u^2} (T_x + T_y)}{\sqrt{4 T_x T_y + u^2 (T_x-T_y)^2} \delta t} \frac{1}{|{\cal W}|^3} + \frac{3 u \sqrt{1 - u^2} (T_y - T_x)}{2 \sqrt{4 T_x T_y + u^2 (T_x-T_y)^2} \delta t} \frac{1}{{\cal W}^4} + o\left(\frac{1}{{\cal W}^4}\right) \,.
\end{equation}  
The first term in this expansion is our eq. \eqref{as1} presented in the main text, while the second term defines the sub-dominant contribution. Interestingly enough,  the amplitude of the second term vanishes in equilibrium conditions (i.e., for $T_x = T_y$) such that in equilibrium the sub-dominant contribution will vanish with ${\cal W}$ at a faster rate.

Because of the algebraic tail, $P({\cal W})$ has only the first moment. In virtue of eq. \eqref{def2}, we have
\begin{equation}
\begin{split}
\overline{\cal W} &= u \overline{\left(\frac{X_t^2 - Y_t^2}{X_t^2 + Y_t^2}\right)}  = u \int^{\infty}_{-\infty} \int^{\infty}_{-\infty} dX_t dY_t \, P(X_t,Y_t)  \, \left(\frac{X_t^2 - Y_t^2}{X_t^2 + Y_t^2}\right) \\
&= \frac{u \sqrt{1-u^2} (T_x - T_y)}{\sqrt{4 T_xT_y + u^2 (T_x - T_y)^2}} \,.
\end{split}
\end{equation}
This is a well-known result which shows that $\overline{\cal W}$ is not equal to zero when simultaneously $u \neq 0$ and $T_x \neq T_y$.
In general, $\overline{\cal W}$ is a non-monotonic function of the coupling constant $u$, 
which vanishes when either $u=0$ or $|u| = 1$ and, hence, attains a maximal value for some intermediate coupling. 

\subsubsection{4. Continuous-time limit $\delta t \to 0$. Asymptotic behavior of $P({\cal W}) $.}

We focus on the PDF $P({\cal W}) $ defined in eq. \eqref{probW} and suppose that ${\cal W}$ is finite. Since $\Lambda(\theta)$ diverges in the limit $\delta t \to 0$, in the leading in $\delta t$ order, we are entitled to 
${\cal W}$-dependent terms in the denominator. Then, using eqs. \eqref{exp} and \eqref{expr}, we have
\begin{equation}
\begin{split}
P({\cal W}) &\simeq  \frac{d^2 \sqrt{\delta t}}{16 \pi} \int^{2 \pi}_0 \frac{d\theta}{(T_y \cos^2(\theta) + T_x \sin^2(\theta))^{1/2} (b \cos^2(\theta) + a \sin^2(\theta) + c \cos(\theta) \sin(\theta))^{3/2} } \\
&= \frac{(4 T_x T_y + u^2 (T_x - T_y)^2)}{8 \pi (T_x + T_y)^2} \sqrt{2 (1 - u^2) \delta t} \int^{2 \pi}_0 \frac{d\phi}{\sqrt{(1 - \Delta \cos(\phi)) (1 + u \sin(\phi) - (1 - u^2) \Delta \cos(\phi))^{3}}} \,, 
\end{split}
\end{equation}
i.e., our eq. \eqref{PWas} with $g(u,\Delta)$ given by
\begin{equation}
\label{g}
g(u,\Delta) = \int^{2 \pi}_0 \frac{d\phi}{\sqrt{(1 - \Delta \cos(\phi)) (1 + u \sin(\phi) - (1 - u^2) \Delta \cos(\phi))^{3}}} \,.
\end{equation}
Likewise $f(u,\Delta)$ in eq. \eqref{f}, $g(u,\Delta)$ diverges logarithmically when $\Delta \to \pm 1$, i.e., either of the temperatures vanishes. Therefore,
 the asymptotic form in eq. \eqref{PWas} is valid only when both temperatures are bounded away from zero. 

\subsubsection{5. Probability density function $P({\cal W})$ in the decoupled case $u = 0$.}

Lastly, we aim to find an analogue of eq. \eqref{u00} which describes the probability density function $P({\cal L})$ 
in the decoupled case $u = 0$. 
For $u = 0$, the probability density function in eq. \eqref{probW} becomes (in the limit $t \to \infty$)
\begin{equation}
\label{lim}
P\left({\cal W}\right) = \frac{1}{2 \pi}  \sqrt{\frac{\delta t}{T_x T_y}} \int^{2 \pi}_0 \frac{\left(T_y \cos^2(\theta) + T_x \sin^2(\theta)\right) d\theta}{\left(\delta t \, {\cal W}^2 + \dfrac{2}{T_x T_y} \left(T_y \cos^2(\theta) + T_x \sin^2(\theta)\right)^2  \right)^{3/2}} \,.
\end{equation} 
Using the first of eqs. \eqref{expr} and the integral identity
\begin{equation}
\frac{p}{\left(a^2 + p^2\right)^{3/2}} = \int^{\infty}_0 \tau \, d\tau \, J_0\left(a \tau\right) \, e^{- p \tau} \,,
\end{equation}
we cast eq. \eqref{lim} into the form
\begin{equation}
\begin{split}
\label{lim1}
P\left({\cal W}\right) &= \sqrt{\frac{\delta t}{2}} \int^{\infty}_0  \tau \, d\tau \, J_0\left(\sqrt{\delta t} \, {\cal W} \tau\right) \,  J_0\left(i \frac{(T_x+T_y)}{\sqrt{2 T_x T_y}}  \Delta \tau\right) \, \exp\left(- \frac{(T_x + T_y)}{\sqrt{2 T_x T_y}} \tau\right) \\
&  = - \sqrt{\frac{\delta t}{2}} \, \frac{d}{d\alpha} \int^{\infty}_0  d\tau \, J_0\left(\sqrt{\delta t} \, {\cal W} \tau\right) \,  J_0\left(i \beta \tau\right) \, \exp\left(- \alpha \tau\right) \,, 
\quad \alpha = \frac{(T_x + T_y)}{\sqrt{2 T_x T_y}} \,, \quad \beta = \frac{(T_x-T_y)}{\sqrt{2 T_x T_y}}  \,.
\end{split}
\end{equation}
The integral in the last line in eq. \eqref{lim1} can be performed explicitly to give
\begin{equation}
\begin{split}
\label{z3}
P\left({\cal W}\right) &=  - \frac{\delta t^{1/4}}{\pi \sqrt{2 i \beta {\cal W}}} \, \frac{d}{d\alpha} \, Q_{-1/2}\left(\frac{\alpha^2 + \delta t \, {\cal W}^2 - \beta^2}{2 i \beta {\cal W} \sqrt{\delta t}}\right) \,,
\end{split}
\end{equation}
where $Q_{-1/2}$ is the Legendre function. It might be also convenient to express the result in eq. \eqref{z3} in an explicit form  using the Gauss hypergeometric functions:
\begin{equation}
\begin{split}
P\left({\cal W}\right) &= \frac{(T_x + T_y) \sqrt{T_x T_y \delta t \, (2 + \delta t \, {\cal W}^2)}}{2 (2 T_y + T_x \delta t \, {\cal W}^2) (2 T_x + T_y \delta t \, {\cal W}^2)} \,_2F_1\left(3/4, 1/4; 1; - \frac{2 (T_x - T_y) \delta t \, {\cal W}^2}{T_x T_y (2 + \delta t \, {\cal W}^2)^2}\right) \\
&+ \frac{(T_x - T_y)^2 \delta t^{3/2} {\cal W}^2}{4 (T_x T_y)^{3/4} (2 + \delta t \, {\cal W}^2)^{3/2} (2 T_y + T_x \delta t \, {\cal W}^2) (2 T_x + T_y \delta t \, {\cal W}^2)} \,_2F_1\left(5/4, 3/4; 2; - \frac{2 (T_x - T_y) \delta t \, {\cal W}^2}{T_x T_y (2 + \delta t \, {\cal W}^2)^2}\right)  \,.
\end{split}
\end{equation}
Importantly, when two processes \eqref{a} are decoupled, the probability density function $P\left({\cal W}\right)$ is an even function of ${\cal W}$, such that it is symmetric around ${\cal W} = 0$, likewise the probability density function of the angular momentum (see eq. \eqref{u00}). The dependence on temperatures and the functional form of the probability density function $P({\cal W})$ are evidently much more complicated than $P({\cal L})$ in eq. \eqref{u00}.

\subsection{D. Probability density function of the moment of inertia}

We evaluate here the probability density function of the moment of inertia ${\cal I}=X^2_t+Y^2_t$, which is a positive definite random variable. Therefore, it is convenient to focus on its moment-generating function 
\begin{equation}
	\Phi_{\cal I}(\lambda) = \overline{\exp\left(-  \lambda {\cal I}\right)} \,, \quad \lambda \geq 0 \,,
\end{equation}
for which one readily obtains the following exact result
\begin{equation}
	\Phi_{\cal I}(\lambda) = \Big(1 + 4 (a+b) \lambda + 4 d^2 \lambda^2\Big)^{-1/2}  \,,
\end{equation}
where the coefficients $a$, $b$, $c$ and $d$ are defined in eqs. \eqref{aa} and \eqref{p}. 
Inverting the latter expression, we find the probability density function
\begin{equation}
	P\left({\cal I}\right) = \frac{1}{2 d} \exp\left(- \frac{(a+b)}{2 d^2} {\cal I}\right) I_0\left(\frac{\sqrt{(a - b)^2 + c^2}}{2 
	d^2} {\cal I}\right) \,,
\end{equation} 
where $I_0$ is the modified Bessel function. 
In the limit $t \to \infty$, $P\left({\cal I}\right) $ attains the form
\begin{equation}
\label{eq:PdI}
\begin{split}
	P\left({\cal I}\right) &= \sqrt{\frac{1 - u^2}{4 T_x T_y + u^2 (T_x - T_y)^2}}  \exp\left(- \frac{T_x+ T_y}{4 T_x T_y + u^2 (T_x - T_y)^2} {\cal I}\right)\\ &\times I_0\left(\frac{\sqrt{4 u^2 T_x T_y + (1 - u^2 + u^4) (T_x - T_y)^2}}{4 T_x T_y + u^2 (T_x - T_y)^2} {\cal I}\right)  \,.
	\end{split}
\end{equation}
In the case of two decoupled  Ornstein-Uhlenbeck processes (i.e., for $u =0$), the probability density function in eq. \eqref{eq:PdI}
simplifies considerably to give \begin{equation}
		P\left({\cal I}\right) = \sqrt{\frac{1}{4 T_x T_y } } \exp\left(- \frac{T_x+ T_y}{4 T_x T_y } {\cal I}\right) I_0\left(\frac{|T_x - T_y|}{4 T_x T_y } {\cal I}\right) \,.
\end{equation}
\begin{figure}
	\centering
	\includegraphics[width=0.3\columnwidth]{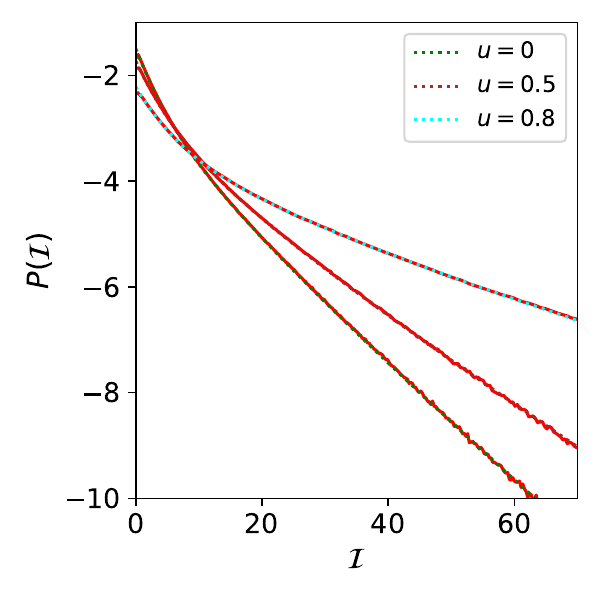}
		\caption{The moment of inertia ${\cal I}$. The probability density function $P({\cal I})$ versus ${\cal I}$ for $T_x = 1$ and $T_y = 5$.  
		 Noisy curves depict the results of numerical simulations, while dot-dashed curves -   the exact expression \eqref{eq:PdI}.
	}  
	\label{fig:4}
\end{figure}

In Fig. \ref{fig:4} the probability density function $P({\cal I})$ is depicted as  function of ${\cal I}$ for three values of $u$ ($u=0$, $0.5$ and $0.8$) and temperatures $T_x=1$ and $T_y=5$. Noisy curves are the solutions of stochastic eqs. \eqref{a} using the Euler-Maruyama algorithm, while the dot-dashed curves
correspond to the analytical expression \eqref{eq:PdI}. One observes a perfect agreement between the numerical
 results and the theoretical prediction. We also note that the most probable value of the moment of inertia ${\cal  I}$
  is attained at ${\cal I} = 0$, which explains why large values of ${\cal W} $ are abundant and therefore, why $P({\cal W})$ possesses heavy power-law tails.

\subsection{E. Numerical simulations.}

\begin{figure}[ht]\label{FIGSM5}
		\centering
	\includegraphics[scale=0.7]{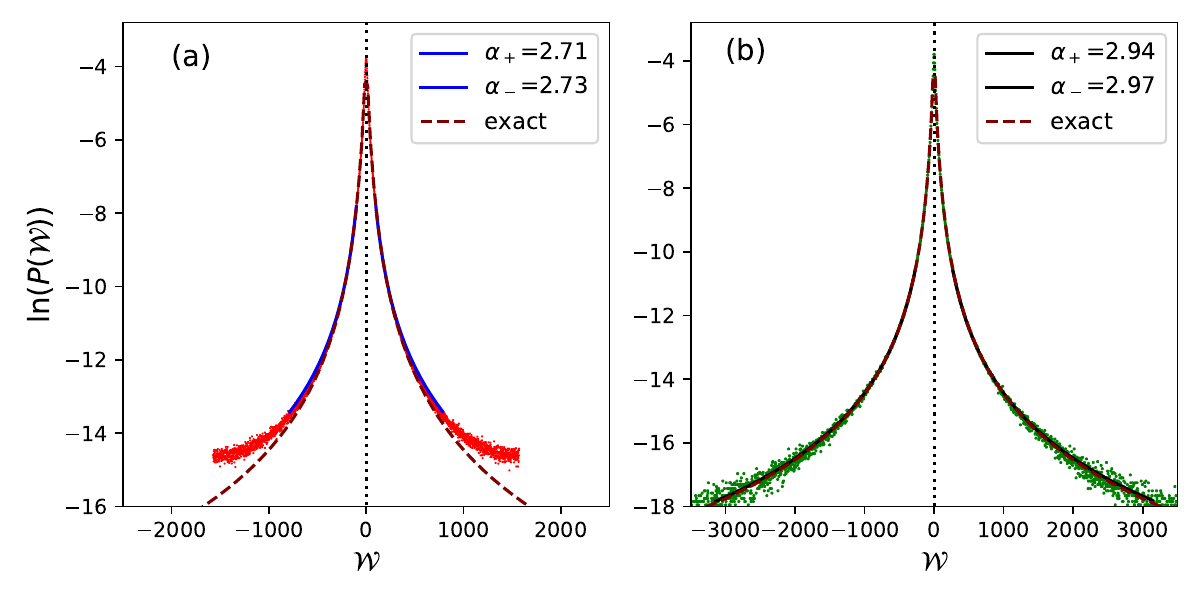}
	\caption{Fits of the results of numerical simulations. Logarithm of the probability density function $P(\cal W)$ versus ${\cal W}$  for $T_x = 1$, $T_y = 5$, $u=1/2$ and $\delta t = 0.002$. The dashed curve is the exact analytical solution in eq. \eqref{probW}. Panel (a): The red noisy curve represents $\ln P(\cal W)$ evaluated using approach I. The blue solid curves are the fits 
	of the red curve by power-law functions $1/|{\cal W}|^{\alpha_-}$ for negative ${\cal W}$ and $1/{\cal W}^{\alpha_+}$ for positive ${\cal W}$. Panel (b): 
	Symbols (filled green circles) depict the numerical results obtained via approach II. The solid black curves are the  fits of the numerical data obtained via approach II by the above power-law functions. The inset shows the best-fit values of the exponents. 
	}
\end{figure}

Numerical simulations of the BG model are performed by using the standard Euler-Maruyama algorithm. 
To compute the angular momentum and the angular velocity, we take advantage of two alternative approaches: \\
-- Approach I. The first approach  
relies on the formal definition of ${\cal W}$ as the rate of change of the 
angular position with respect to time
\begin{equation}
\label{ww}
{\cal W}=\frac{\delta \theta}{\delta t} \,,
\end{equation}
where the instantaneous value of the polar angle is expressed via the cartesian coordinates as
\begin{equation}
\theta = {\rm atan2}(Y_t,X_t)
\end{equation}
where 
the function ${\rm atan2}$ is the $2$-argument arctangent \cite{arc}.
While determining ${\cal W}$ through eq. \eqref{ww}, we exercise care that at each incremental step $\delta \theta$ does not exceed $\pi$, which is controlled by the increment $\delta t$. 
Note also that within this approach  the distribution of the angular velocity has its support on the interval $[-\pi/\delta t,\pi/\delta t]$, so that while 
evaluating $P({\cal W})$ we consider only such values of ${\cal W}$ that are away from the boundaries, in order to avoid  aliasing effects appearing close to the boundaries of the interval. To access the behavior at larger values of ${\cal W}$, we have to diminish the increment $\delta t$. Of course, this can be readily 
done in numerical simulations, but in experiments the value of $\delta t$ is bounded from below by the maximal frequency at which the images of the trajectories are taken.

Once ${\cal W}$ is determined, the angular momentum is found via the relation
\begin{equation}
{\cal L}=(X_t^2+Y_t^2){\cal W} \,.
\end{equation}
Note that this approach is appropriate 
  for the analysis of discrete-time trajectories recorded both in numerical simulations 
 and  in experiments (see Fig. \ref{F:3} in the main text), because it necessitates the position data only. In 
 Fig. \ref{FIGSM2} the results obtained using the approach I are depicted by noisy red curves.\\
-- Approach II. The second approach hinges on  the standard recursive 
solution of the time-discretized Langevin eqs. \eqref{az}, for which the angular momentum ${\cal L}$ and the angular velocity ${\cal W}$ are determined by eqs. \eqref{Lz} and \eqref{def2}, respectively. 
 Generating in numerical simulations the dimensionless
 noises $\eta_x(t)$ and $\eta_y(t)$, we build recursively the trajectories $X_t$ and $Y_t$ and eventually calculate the angular momentum and the angular velocity. Results obtained via this approach are depicted in Fig.  \ref{FIGSM2}  by noisy green curves. Note that this second approach is only applicable for the numerical simulations.

Lastly, we comment on the discrepancy between the theoretically predicted exponent ($= 3$), characterizing the tails of the probability density function $P({\cal W})$, and somewhat lower values deduced from the fitting of the experimentally-evaluated PDFs. To this end, we resort to numerical simulations of the BG model using approaches I and II and preform the fitting of the numerical data, which we fully control.   
These fits are presented in Fig. S5. We observe that for the same value of $\delta t$ (which is quite small, $\delta t = 0.002$), the PDF $P({\cal W})$ obtained within the approach I is defined on a much shorter interval, as compared to 
the one obtained in terms of approach II, and hence, the large-${\cal W}$ tails are systematically heavier than  
the exact solution. The best-fit values of the exponents $\alpha_{\pm}$ obtained within the approach I are $\alpha_- \approx 2.73$ and $\alpha_+ \approx 2.71$, respectively, which are somewhat higher that the exponents deduced from the experimental data (see Fig. 3 in the main text). This implies that the accuracy of approach I, which is used to analyze the experimental data, can be somewhat improved by reducing $\delta t$, but the latter cannot be made arbitrarily small 
due to natural limits imposed on the sampling frequency of imaging the trajectories in Ref. \cite{argun}.
In turn, approach II appears to be more accurate giving the values of the exponents that are closer to the predicted value. 

%\end{widetext}

\end{document}